# Saturation Throughput Analysis of IEEE 802.11b Wireless Local Area Networks under High Interference considering Capture Effects

Ponnusamy Kumar
Department of Electronics and Communication Engineering
K.S.Rangasamy College of Technology
Tiruchengode, Namakkal, Tamilnadu, India

A.Krishnan
Department of Electronics and Communication Engineering
K.S.Rangasamy College of Technology
Tiruchengode, Namakkal, Tamilnadu, India

*Abstract*— **Distributed contention based Medium Access Control (MAC) protocols are the fundamental components for IEEE 802.11 based Wireless Local Area Networks (WLANs). Contention windows (CW) change dynamically to adapt to the current contention level: Upon each packet collision, a station doubles its CW to reduce further collision of packets. IEEE 802.11 Distributed Coordination Function (DCF) suffers from a common problem in erroneous channel. They cannot distinguish noise lost packets from collision lost packets. In both situations a station does not receive its ACK and doubles the CW to reduce further packet collisions. This increases backoff overhead unnecessarily in addition to the noise lost packets, reduces the throughput significantly. Furthermore, the aggregate throughput of a practical WLAN strongly depends on the channel conditions. In real radio environment, the received signal power at the access point from a station is subjected to deterministic path loss, shadowing and fast multipath fading. In this paper, we propose a new saturation throughput analysis for IEEE 802.11 DCF considering erroneous channel and capture effects. To alleviate the low performance of IEEE 802.11 DCF, we introduce a mechanism that greatly outperforms under noisy environment with low network traffic and compare their performances to the existing standards. We extend the multidimensional Markov chain model initially proposed by Bianchi[3] to characterize the behavior of DCF in order to account both real channel conditions and capture effects, especially in a high interference radio environment.**

## I. INTRODUCTION

The use of IEEE 802.11 wireless local area networks (WLANs) has been spreading quickly. One of the channel access mechanisms in IEEE 802.11 is Distributed Coordination Function (DCF). DCF is based on Carrier Sense Multiple Access with Collision Avoidance(CSMA/CA) algorithm. In this mechanism, a station waits for a quiet period in wireless media, and then begins to transmit data while detecting collisions. The time lapse between successive carrier senses, when channel is occupied, is given by a back-off counter which has an initial random value within a predetermined range. The standard also defines an optional access method, PCF, which is for time bounded traffic. Since the DCF mechanism has been widely adopted in wireless networks, we focus our analysis only on this mechanism.

Many research efforts have been done to study the IEEE 802.11 DCF performance, by both of analysis and simulation. Most of them assume the ideal channel condition, which means that the packet corruptions are only due to collisions[3-5]. A detailed analysis for the multi-access behavior in the 802.11 DCF is presented in [6], where the packet sending probability p, which depends on different contention window size, is computed by approximating the 802.11 DCF under saturated traffic as a p-persistent CSMA protocol. This approximation is very useful for the analysis of the DCF and several papers [7-10] have adopted this approach and analyzed saturated DCF performance. Bianchi [3] suggests a Markov model to represent the exponential backoff process, and Wu et al. [8] use the same model and take the packet retry limit into account.

In [11] , based on the IEEE 802.11 DCF, a novel scheme named DCF is proposed to improve the performance of Wireless Local Area Network (WLAN) in fading channel. Impact of bursty error rates on the performance of wireless local area network is studied in [12]. The throughput and delay were analyzed in ideal and error-prone channels.

In [13], the authors proposed a fast collision resolution (FCR) algorithm. In this algorithm, when a station detects a busy period, it exponentially increases its contention window and generates a new backoff counter. In case that a station detects a number of consecutive idle slots, it exponentially reduces the backoff counter. In [14], the authors proposed a new backoff algorithm to measure the saturation throughput under several conditions and several set of parameters which are adjusted dynamically according to the network conditions.

In [15], the authors presented an extension of Bianchi's model to a non saturated environment. They modified the multi-dimensional Markovian state transition model by including state, characterizing the system when there are no packets to be transmitted in the buffer of a station. These states are called post backoff states and denote a kind of virtual backoff counter initiated prior to packet arrival. In [16], the authors propose a new scheme for IEEE 802.11 DCF to alleviate the low performance of the high date rate stations for asynchronous networks. They introduce an adaptive mechanism to adjust the packet size according to the data rate, in which the stations occupy the channel equal amount of time.

In real radio environment, the signal power at access point from a given station will be subjected to deterministic path loss, shadowing and fast multi-path fading. Due to this, when more than one station simultaneously transmit to access point, the channel is successfully captured by a station whose signal power level is stronger than the other stations and thus increases the actual throughput. This phenomenon is called capture effect. In [17], the authors presented a Markov model to analyze the throughput of IEEE802.11 considering transmission errors and capture effects over Rayleigh fading channels in saturated network conditions. Their model is very accurate when the contention level of a network is high. In [18], we have presented a novel scheme for DCF under non saturated traffic condition. In [19], we extend [3] and presented the non saturation throughput analysis for heterogeneous traffic. Hadzi-velkov and Spasenovski [20] have investigated the impact of capture effect on IEEE 802.11 basic service set under the influence of Rayleigh fading and near/far effect.

Liaw et al. [22] introduced an idle state, not present in Bianchi's model [3], accounting for the case in which the station buffer is empty after a successful completion of a packet transmission. The probability that there is at least a packet in the buffer after a successful transmission is assumed to be constant and independent of the access delay of the transmitted packet. In [23], we presented the performance study for multihop network in string topology.

In this paper, we present an analytical model to study the saturation behavior of the IEEE802.11 DCF considering erroneous channel and capture effects. We differentiate channel induced errors from packet collision in order to optimize the performance of CSMA/CA under the saturated network condition.

The rest of the paper is organized as follows: Section II describes our model for basic access mechanism under saturated traffic condition. Performance of the proposal scheme is analyzed in section III. Finally, section VI concludes this paper.

## II. Performance Analysis For 802.11 DCF In Saturated Trafic Condition

In this section, we present a discrete time bi-dimensional Markov model for evaluating the saturation throughput of the DCF under non ideal channel conditions considering capture effects. Saturated traffic condition means that all users always have a packet available for transmission. Throughput under saturated traffic situation is the upper limit of the throughput achieved by the system, and it represents the maximum load the system can carry in the stable condition.

Let process s(t) be the stochastic process representing the backoff stage of a given station at the given time t. A second process b(t) is defined, representing backoff time counter of the station. Backoff time counter is decremented at the start of every idle backoff slot. The backoff counter is an integer value uniformly chosen from $[0,W_i-1]$ where $W_i$ denotes the contention window at the $i^{th}$ backoff stage. The backoff stage 'i' is incremented by one for each failed transmission attempt up to the maximum value m, while the contention window is doubled for each backoff stage up to the maximum value $W_{max} = 2^m W_{min}$.

Letting $W_{min} = W_0$, we can summarize the W as,

$$W_i = \begin{cases} 2^i W_0, \ 0 \le i \le m \\ 2^m W_0, \ i > m \end{cases} \quad (1)$$

The main aim in this section is to modify the MAC protocol in order to enhance the performance of MAC protocol in the event of channel induced errors. The assumption that all frame losses are due to collisions between WLAN devices is generally not true in a noisy wireless environment. However, unsuccessful reception of the data frame can also be caused by channel noise or other interference.

In case of unsuccessful transmission the basic BEB mechanism will double the contention window size by considering channel errors as a packet collision. This process will unnecessarily increase the backoff overhead and intern increases channel idle slots. In order to alleviate this problem we propose a new mechanism that takes advantage of a new capability to differentiate the losses, and thereby sharpen the accuracy of the contention resolution process. When the frame is corrupted by the channel induced noise, we maintain the same contention size instead of doubling it.

### *A. Loss differentiation method for basic access mechanism*

Basic access mechanism is the default access method in DCF and employs a two-way handshaking procedure. The loss differentiation for basic access is not straightforward because it provides only ACK feedback from receivers. The loss differentiation method for WLAN has been proposed in [24]. The following describes a loss differentiation method for basic access which requires minimum modifications to the legacy standard to provide additional feedback. The data frame can be functionally partitioned into two parts: header and body. The MAC header contains information such as frame type, source address and destination address, and comes before the MAC body, which contains the data payload.

In a WLAN with multiple stations sharing a common channel, a collision occurs when two or more stations starts transmission in the same time slot, which will likely corrupt the whole frame (header plus body) at the receiver end. On the other hand, a frame transmission that is not affected by collision with transmission from another WLAN station may still be corrupted by noise and interference. However, under the condition that the signal-to-noise-plus-interference ratio (SINR) is reasonable to maintain a connection between the sending and receiving stations, the receiver is likely able to acquire the whole data frame and decode it, as the physical header is transmitted at the base data rate for robustness (e.g., in 802.11b, the 192-bit physical header is always transmitted at 1Mbps). In this case, the noise or interference may result in a few bit-errors that cause a Frame Check Sequence (FCS) error in the decoded data frame, which is then discarded by the receiver station. As the MAC header (18-30 bytes) is typically much shorter than the MAC body (e.g., a typical

Internet Protocol datagram is several hundreds to a couple of thousands bytes long), when FCS fails, it is much more likely caused by bit errors in the body than the header.

If the MAC header is correctly received but the body is corrupted, the receiver can observe the MAC header content to learn the identity of the sender and to verify that it is the intended receiver. To verify the correctness of the MAC header, a short Header Error Check (HEC) field can be added at the end of the header as shown in Fig.1 in order to provide error checking over the header, while the FCS at the end of the frame provides error checking over the entire MAC frame. Note that the use of HEC in the header is not a new concept as it has been adopted in many other communication systems, such as asynchronous transfer mode and Bluetooth, all of which includes a 1-byte HEC or header check sequence (HCS) field in their header. With the HEC, when a data frame is received and FCS fails, the HEC can be verified to see if the header is free of error, and if so, proper feedback can be returned to the sender identified by the MAC header.

As discussed above, FCS failure but correct HEC in a frame reception is a good indication that the frame has been corrupted by transmission errors rather than a collision. Because in the basic access mechanism, only ACK frames are available to provide positive feedback, a new control frame NAK needs to be introduced to inform the sender that the data frame transmission has failed and the failure is due to transmission errors; i.e., the data frame has suffered a transmission loss. On the other hand, if the sender receives neither a NAK nor an ACK after sending a data frame, it is a good indication that the frame transmission has suffered a collision loss. The NAK frame can be implemented with exactly the same structure as the ACK frame except for a one-bit difference in the frame type field in the header, and is sent at the same data rate as an ACK frame. The transmission of a NAK does not consume more bandwidth or collide with other frames because it is transmitted SIFS after the data frame transmission and occupies the time that would have been used by the transmission of an ACK.

The HEC field is a necessary modification to the standard because without it, when the FCS fails, the receiver would not be able to determine if the header is in error and would not be able to trust the sender address in the header for returning the NAK. The HEC field (which can be 1 or 2 bytes) costs an extra overhead. But it can be calculated that the overhead due to the extra field to the total transmission time is much less than 1%. Therefore the overhead is negligible. Comparing the two loss differentiation methods, RTS/CTS access is useful when the data frame size or the number of stations is very large or there are hidden terminals. However, as it consumes extra time for RTS/CTS exchange, RTS/CTS access is less efficient than basic access in other cases.

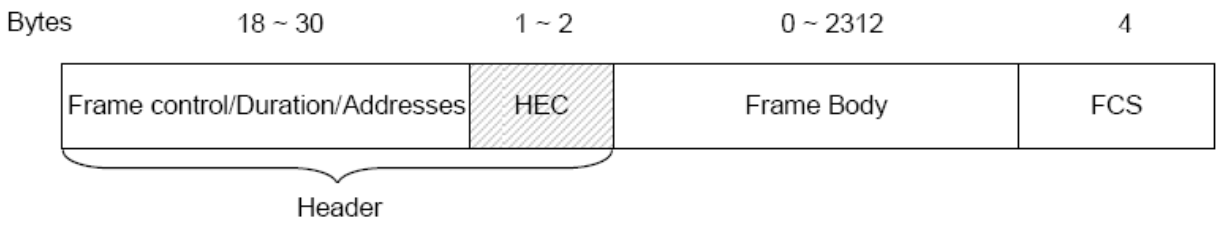

Figure 1. Frame format for Basic access mechanism

The loss differentiation method for RTS/CTS access does not involve any modification to the standard. The loss differentiation method for basic access, however, needs two minor modifications to the current standard: the HEC field and the NAK frame, which are both easy to implement.

### B. *Analitical modeling of new backoff algorithm*

Based on the above consideration, let us discuss the Markovian model shown in Fig.2, assuming saturation network conditions. We assume that each station has m+1 stages of backoff process. The value of the backoff counter is uniformly chosen in the range (0,$W_i$-1), where $W_i = 2^i W_{min}$ and depend on the station's backoff stage *i.* A station in (i,0) state will transit into (i+1,k) state in the event of collision without capture effect. On the other hand, the model transits from (i,0) to (0,k) state if frame is successfully captured. From state (i,0), the station re-enters the same backoff stage (i,k) in case of unsuccessful transmission due to transmission errors.

The main approximation in our model is that, at each transmission attempt, each packet collides with constant and independent probability $P_{col}$ regardless of previously suffered attempts and transmission errors occur with probability $P_e$ due to the erroneous channel. We also assume that the channel is captured by a station with the probability $P_{cap}$ in the event of collision. Based on the above assumptions we can derive the transition probabilities:

$$P\{i,k/i,k+1\} = 1,\ \ k \in [0, W_i - 2],\ \ i \in [0,m]$$

$$P\{0,k/i,0\} = ((1-P_{col})(1-P_e) + P_{col}P_{cap})/W_0,\ \ k \in [0,W_i - 1],\ \ i \in [0,m]$$

$$P\{i,k/i,0\} = (1-P_{col})P_e/W_0,\ \ k \in [0,W_i - 1],\ \ i \in [0,m]$$

$$P\{i,k/i-1,0\} = (1-P_{cap})P_{col}/W_i,\ \ k \in [0,W_i - 1],\ \ i \in [1,m]$$

$$P\{m,k/m,0\} = ((1-P_{cap})P_{col} + (1-P_{col})P_e)/W_m,\ \ k \in [0,W_m - 1] \qquad (2)$$

The first equation represents that, at the beginning of each time slot, the backoff time is decremented. The second equation states that, the initialization of backoff window after successful transmission for a new packet. The third equation accounts that, the maintenance of backoff window in the same stage, if channel error is detected. The fourth and fifth equations represent that, the rescheduling of backoff stage after unsuccessful transmission.

Let the stationary distribution of the chain be $b_{i,k} = \lim_{t \to \infty} P\{s(t)=i, b(t)=k\}, i \in (0,m), k \in (0, W_{i-1})$ . To obtain the closed form solution we first consider the following relations:

$$b_{i,0} = b_{i-1,0}\{P_{col}(1-P_{cap})\} + b_{i,0}\{P_e(1-P_{col})\}$$

$$= \left(\frac{P_{col}(1-P_{cap})}{1-(1-P_{col})P_e}\right) b_{i-1,0}$$

$$= \left(\frac{P_{col}(1-P_{cap})}{1-(1-P_{col})P_e}\right)^i b_{0,0} \quad 0 < i < m \qquad (3)$$

and,

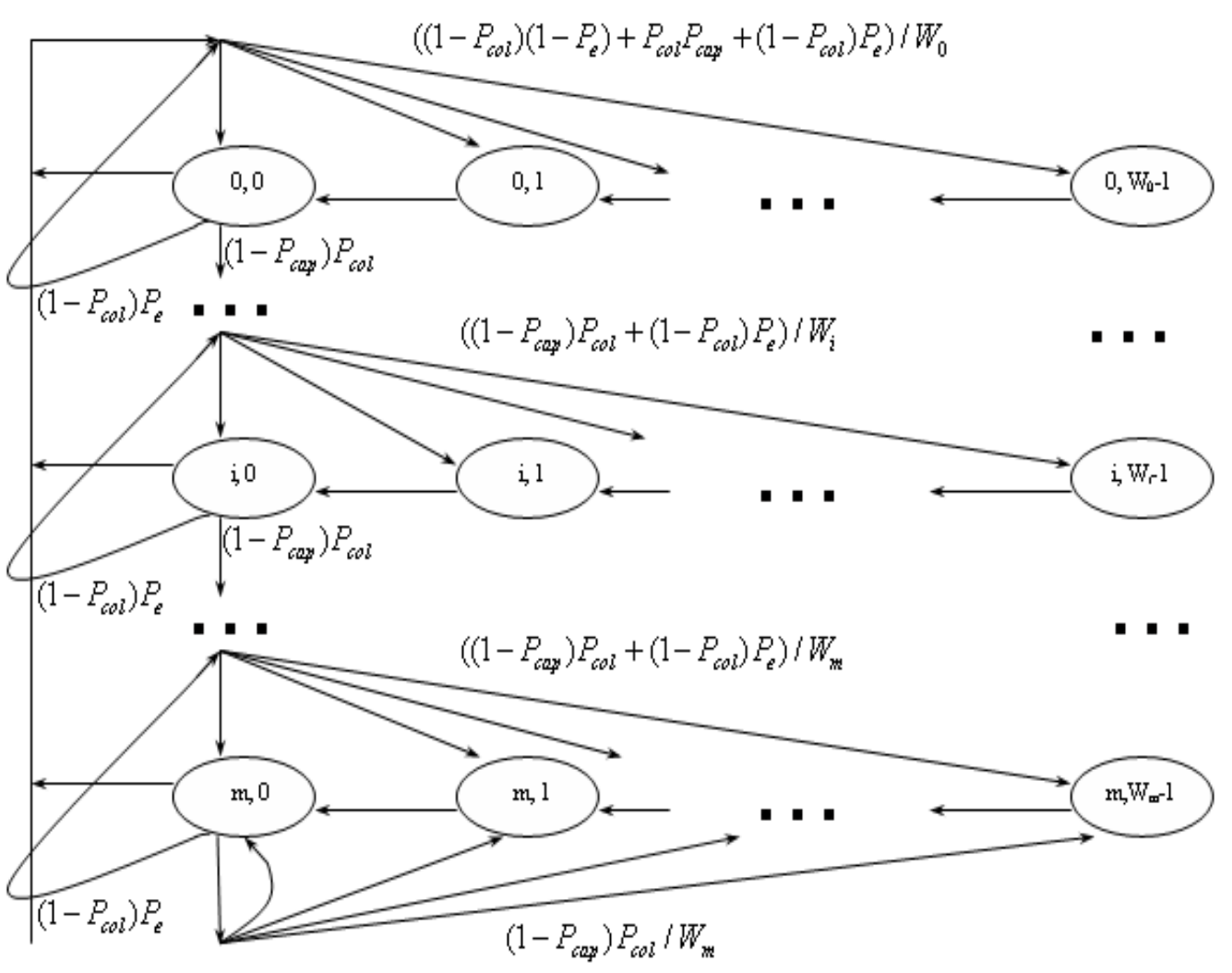

Figure 2. Markov chain model for the backoff procedure of a station

$$b_{m,0}=b_{m-1,0}(1-P_{cap})P_{col}+b_{m,0}P_{col}(1-P_{cap})+b_{m,0}(1-P_{col})P_e \qquad (4)$$

$$b_{m,0}\left\{1-\left(\frac{P_{col}(1-P_{cap})}{1-(1-P_{col})P_e}\right)\right\}=b_{m-1,0}\left(\frac{P_{col}(1-P_{cap})}{1-(1-P_{col})P_e}\right) \qquad (5)$$

from which we obtain the following relation,

$$b_{m,0}=\frac{\left(\frac{P_{col}(1-P_{cap})}{1-(1-P_{col})P_e}\right)^m b_{0,0}}{1-\left(\frac{P_{col}(1-P_{cap})}{1-(1-P_{col})P_e}\right)} \qquad (6)$$

A closed-form solution to the Markov chain owing to the chain regularities, for each k ϵ (1,$W_i$-1), shown as:

$$b_{i,k}=\frac{W_{i-k}}{W_i}\begin{cases}((1-P_{col})(1-P_e)+P_{col}P_{cap})\sum_{i=0}^{m}b_{i,0}+b_{i,0}(1-P_{col})P_e, & i=0\\ P_{col}(1-P_{cap})b_{i-1,0}+(1-P_{col})P_e b_{i,0}, & 1\le i\le m\\ (P_{col}(1-P_{cap}))(b_{m-1.0}+b_{m,0})+(1-P_{col})P_e b_{m,0}, & i=m\end{cases} \qquad (7)$$

By means of relations (3), (5) and remembering

$$\sum_{i=0}^{m}b_{i,0}\left(1-\frac{P_{col}(1-P_{cap})}{1-(1-P_{col})P_e}\right)=b_{0,0}$$

we rewrite the relation (7) as:

$$b_{i,k}=\frac{W_i-k}{W_i}b_{i,0} \quad i\in(0,m),\ k\in(0,W_i-1) \qquad (8)$$

Thus, by relations (3), (5) and (8), all the values of $b_{i,k}$ are expressed as a function of $b_{0,0}$. Considering normalization conditions, and making use of the above equations we obtain the following relation:

$$\sum_{i=0}^{m}\sum_{k=0}^{Wi-1}b_{i,k}=1$$

$$=\frac{b_{0,0}}{2}\left[W_0\left(\sum_{i=0}^{m-1}(2P_t)^i+\frac{(2P_t)^m}{1-P_t}\right)+\frac{1}{1-P_t}\right]$$

from which, we obtain,

$$b_{0,0}=\frac{2(1-P_t)(1-2P_t)}{W_0(1-P_t)(1-(2P_t)^m)+W_0(2P_t)^m(1-2P_t)+(1-2P_t)} \qquad (9)$$

where we assume,

$$P_t=\frac{P_{col}(1-P_{cap})}{1-(1-P_{col})P_e}$$

Assuming error free channel with no capture effects, i.e., $P_e = P_{cap} = 0$, then (9 ) can be rewritten as,

$$b_{0,0}=\frac{2(1-2P_{col})(1-P_{col})}{(W+1)(1-2P_{col})+W(1-(2P_{col})^m)P_{col}} \qquad (10)$$

which is similar to $b_{0,0}$ found in Bianchi's model[3] under saturated load conditions.

Now we can express the probability $\tau$ that a station transmits in a randomly chosen slot time when the backoff time is zero as,

$$\tau=\sum_{i=0}^{m}b_{i,0}=\frac{b_{0,0}}{1-P_t} \qquad (11)$$

By substituting (9) in (11), we obtain the following relation.

$$\tau=\frac{2(1-2P_t)}{W_0(1-P_t)(1-(2P_t)^m)+W_0(2P_t)^m(1-2P_t)+(1-2P_t)} \qquad (12)$$

Note that, when m=0, that is no exponential backoff is considered, and assuming $P_{cap}=P_e=0$, the probability $\tau$ results to be independent of collision probability under saturated traffic condition

$$\tau=\frac{2}{W_0+1} \qquad (13)$$

which is the result found in[3] for constant backoff window.

However, in general, the probability τ depends on the conditional collision probability $P_{col}$, capture probability $P_{cap}$ and probability of packet loss $P_e$. In our model we assume basic access method to compute the conditional collision probability $P_{col}$. To determine the value $P_{col}$ it is sufficient to note that the probability that a transmitted packet encounters a collision if in a given time slot, at least one of the remaining (n-1) stations transmits another packet simultaneously. The conditional collision probability also depends on the capture probability because capture effect is the sub event of collision, i.e. without collision there is no capture effect. Therefore the probability $P_{col}$ can be expressed as,

$$P_{col}=1-(1-\tau)^{n-1}-P_{cap} \qquad (14)$$

Our proposed model considers deterministic power loss and multipath fast fading of transmitted signals into account. We also assume that there is no direct path between transmitter and receiver within Basic Service Set(BSS), which means the envelop of received signal is Rayleigh faded. To compute the capture probability, we use the model proposed by Hadzi-velkov and Spasenovski[10]. In Rayleigh fading channel, the transmitted instantaneous power is exponentially distributed according to

$$f(p) = \frac{1}{p_0}\exp(-\frac{p}{p_0}), \qquad p > 0 \tag{15}$$

where $p_0$ represent the local mean power of the transmitted frame at the receiver and is determined by

$$p_0 = A.r_i^{-x}.p_t$$

where $r_i$ is the mutual distance from transmitter to receiver, x is the path loss exponent, $A.r_i^{-x}$ is the deterministic path loss and $p_t$ is the transmitted signal power. The path loss exponent for indoor channels in picocells is typically taken as 4. During simultaneous transmission of multiple stations, a receiver captures a frame if the power of detected frame $p_d$ sufficiently exceeds the joint power of 'n' interfering contenders

$$p_{\text{int}} = \sum_{k=1}^{n} p_k$$

by a certain threshold factor for the duration of a certain time period. Thus capture probability is the probability of signal to interference ratio

$$\gamma = \frac{p_d}{p_{\text{int}}} \tag{16}$$

exceeding the product $z_0 g(S_f)$ where $z_0$ is known as the capture ratio and $g(S_f)$ is processing gain of the correlation receiver. The processing gain introduces a reduction of interference power by a factor $g(S_f)$, which is inversely proportional to spreading factor $S_f$. The conditional capture probability $P_{cap}$ can be expressed over i interfering frames as,

$$\begin{aligned} P_{cap}(z_0 g(S_f) \mid i) &= prob(\gamma > z_0 g(S_f)/i) \\ &= [1 + z_0 g(S_f)]^{-i} \end{aligned} \tag{17}$$

For Direct Sequence Spread Spectrum(DSSS) using 11 chip spreading factor ($s_f$=11),

$$g(S_f) = \frac{2}{3S_f}$$

Now the frame capture probability can be expressed as,

$$P_{cap}(z_0, n) = \sum_{i=1}^{n-1} R_i P_{cap}(z_0 g(S_f) \mid i) \tag{18}$$

where $R_i$ is the probability of 'i' interfering frames being generated in the generic time slot, according to

$$R_i = \binom{n}{i+1} \tau^{i+1} (1-\tau)^{n-i-1} \tag{19}$$

Next step is the computation of the saturation system throughput, defined as the fraction of time the channel is used successfully to transmit the bits. Let $P_{tr}$ be the probability that there is at least one transmission in the considered time slot, with n stations contending for the channel, and each transmits with probability τ,

$$P_{tr} = 1 - (1-\tau)^n \tag{20}$$

The probability $P_s$ that a transmission on the channel is successful is given by the probability that exactly one station transmit on the channel or probability that two or more stations transmit simultaneously where one station captures the channel due to capture effects,

$$P_s = \frac{n\tau(1-\tau)^{n-1} + P_{cap}}{1-(1-\tau)^n} \tag{21}$$

Now we can express throughput as,

$$S = \frac{E[payload\ transmitted\ in\ a\ timeslot]}{E[length of\ a\ timeslot]}$$

$$= \frac{P_{tr}P_s(1-P_e)E[PL]}{(1-P_{tr})\sigma + P_{tr}(1-P_s)T_c + P_{tr}P_sP_eT_e + P_{tr}P_s(1-P_e)T_s} \tag{22}$$

where, $T_c$ is the average time that the channel sensed busy due to collision, $T_s$ is the average time that the channel sensed busy due to successful transmission, $T_e$ is the average time that the channel is occupied with error affected data frame and σ is the empty time slot. For the basic access method we can express the above terms as,

$T_c$= H + E[PL] + $ACK_{timeout}$

$T_s$= H + E[PL] + SIFS + ACK + DIFS + $2\tau_d$

$T_e$= H + E[PL] + NAK

Here, H – Physical header + MAC header

E[PL] – Average payload length and

$\tau_d$ – propagation delay

## III. PERFORMANCE EVALUATIONS

In what follows, we shall present the results for the data rate of 11Mbps. In the results presented below we assume the following values for the contention window: $W_{min}$=32, m=5 and $W_{max}$=1024. The network parameters of 802.11b are given in Table I. We have also examined 802.11b with other possible parameter values. We use the method given in the IEEE standards [2] to calculate the bit error rates (BERs) and frame error rates (FERs) in a WLAN. This method has also been used in [25]- [27] to study WLAN performance. It is briefly described as follows.

TABLE I. NETWORK PARAMETERS

| | |
|---|---|
| MAC header | 24 bytes |
| PHY header | 16 bytes |
| Payload size | 1024 bytes |
| ACK | 14 bytes |
| NAK | 14 bytes |
| Basic rate | 1Mbps |
| Data rate | 11Mbps |
| $\tau_d$ | 1 $\mu s$ |
| Slot time | 20 $\mu s$ |
| SIFS | 10 $\mu s$ |
| DIFS | 50 $\mu s$ |
| ACK timeout | 300 $\mu s$ |

*A. Bit error rate (BER) model for 802.11b*

First, the symbol error rate (SER) is calculated based on the signal-to-noise-plus-interference ratio (SINR) at the receiver. It is assumed that the interference and noise affect the desired signal in a manner equivalent to additive white Gaussian noise (AWGN). Given the number of bits per symbol, the SER is then converted into an effective BER. IEEE 802.11b uses DBPSK modulation for basic data rate at 1Mbps and complementary code keying (CCK) modulation to achieve its higher data rates (5.5 Mbps and 11 Mbps). The SER in CCK [2] has been determined as:

$$SER = \sum Q(\sqrt{2 \times SINR \times R_c \times D_c}) \qquad (23)$$

where, $R_c$ is the code rate, $D_c$ is the codeword distance, and, $\Sigma$ is over all codewords. For 11 Mbps data rate, the SER is given by

$$\begin{aligned} SER_{11Mbps} = {} & 24 \times Q(\sqrt{4 \times SINR}) + 16 \times Q(\sqrt{6 \times SINR}) \\ & + 174(\sqrt{8 \times SINR}) + 16 \times Q(\sqrt{10 \times SINR}) \\ & + 24 \times Q(\sqrt{12 \times SINR}) + Q(\sqrt{4 \times SINR}) \end{aligned} \qquad (24)$$

As each symbol encodes 8 bits in 11 Mbps, the BER is

$$BER_{11Mbps} = \frac{2^{8-1}}{2^8 - 1} \times SER_{11Mbps} = \frac{128}{255} \times SER_{11Mbps} \qquad (25)$$

The SER for 5.5 Mbps is calculated as,

$$SER_{5.5Mbps} = 14 \times Q(\sqrt{8 \times SINR}) + Q(\sqrt{16 \times SINR}) \qquad (26)$$

And

$$BER_{5.5Mbps} = \frac{2^{4-1}}{2^4 - 1} \times SER_{11Mbps} = \frac{8}{15} \times SER_{11Mbps} \qquad (27)$$

The SER in DBPSK modulation scheme has been determined as:

$$SER_{1Mbps} = Q(\sqrt{11 \times SINR}) \qquad (28)$$

For 1Mbps mode, because each symbol encodes a single bit, the BER is the same as SER. In case of 2Mbps, the BER is calculated as,

$$SER_{2Mbps} = Q(\sqrt{5.5 \times SINR}) \qquad (29)$$

When the BERs have been determined, the FERs of the data and control frames are derived from the BERs and the frame lengths.

Now we can drive the frame error rate, combining BER values for both header and payload as:

$$P_e = 1 - (1 - BER_{1Mbps})^{PHY} + (1 - BER_{11Mbps})^{(MAC+DATA)} \qquad (30)$$

where, PHY is the length of the physical header, MAC is the length of the MAC header and DATA is the length of the packet payload.

*B. Numerical results and discussions*

The behavior of the transmission probability ‘τ’ is depicted in Fig.3 for basic access method as a function of SINR. The curves have been drawn for the capture threshold 6dB, number of contenting stations 10 and payload size 1024 bytes. The results shows that for increasing the channel quality, the transmission probability ‘τ’ increases and reaches the steady state, above which the channel is assumed as ideal. The transmission probabilities of our proposed model and model [17] are clearly highlighted in the above results. The Bianchi’s transmission probability is depicted as horizontal lines due to independence of the Bianchi’s model on both capture effects and channel errors.

Fig.4 shows the behavior of the saturation throughput as a function of the number of the contending stations for basic access mechanism. The curves have been drawn for the capture threshold 6dB, SINR 7dB and payload size 1024 bytes. The curves clearly show that, when the network load is moderate our proposed algorithm performs well. On the other hand, throughput can be higher than the model[17] for a low number of contending stations in the considered scenario. When the number of contending stations increases then the achievable throughput will approach the saturation throughput obtained in model[17].

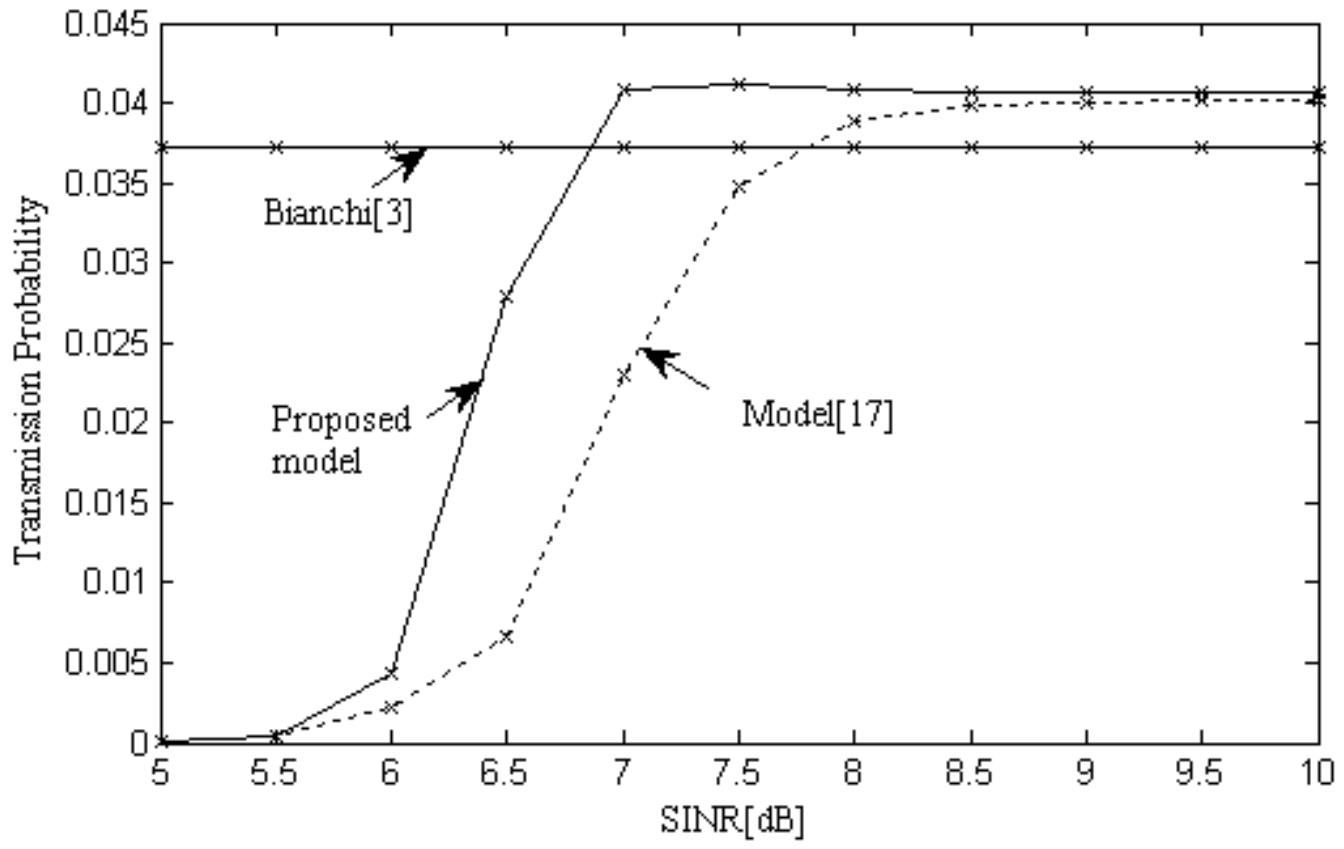

Figure 3. Transmission probability as a function of SINR for basic access mechanism. Curves have been obtained for the capture threshold 6dB, payload size 1024 bytes and number of contending stations 10.

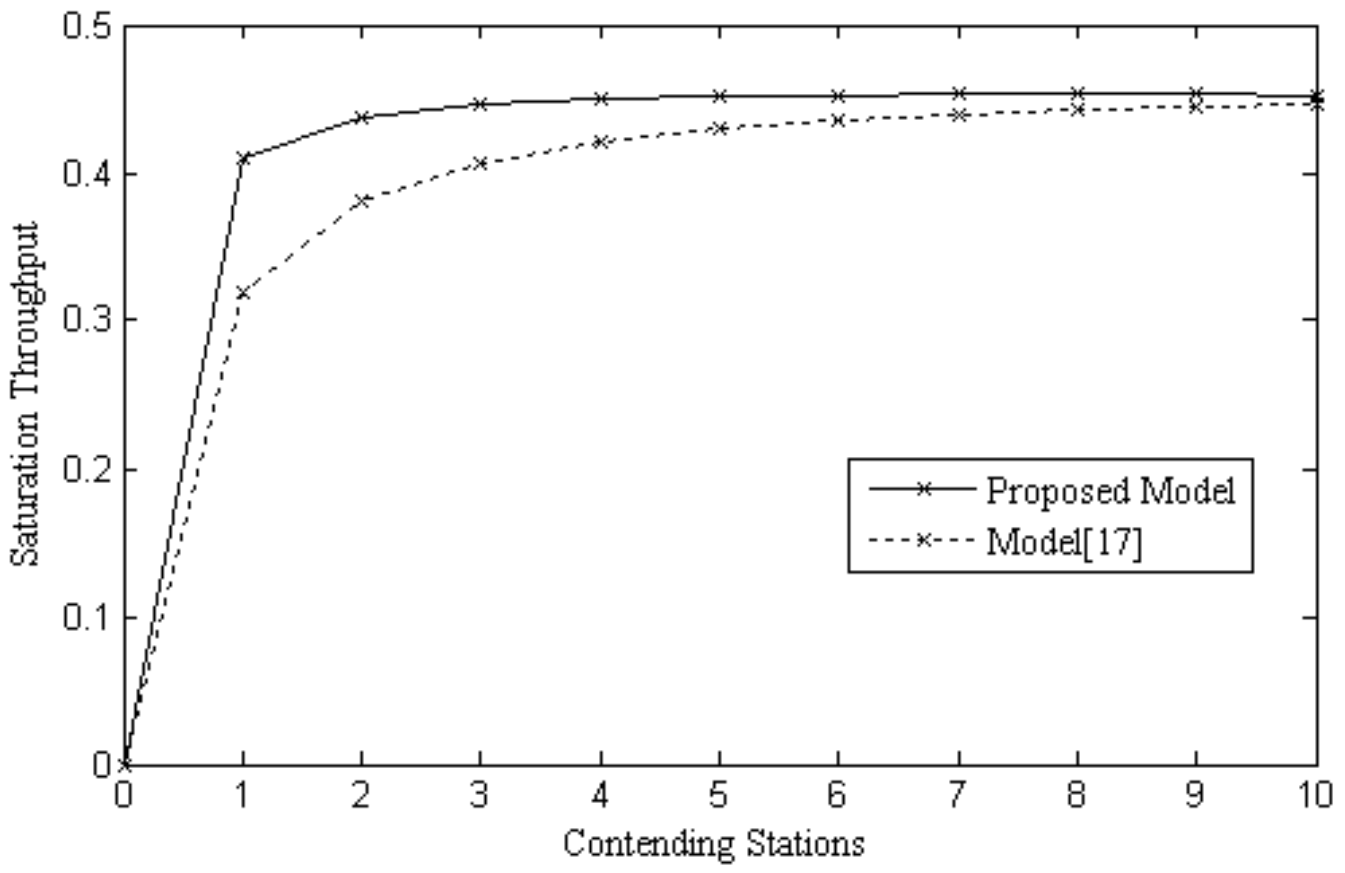

Figure 4. Saturation throughput for basic access mechanism as a function of the number of contending stations for capture threshold 6dB and SINR 7dB while payload size 1024 Bytes.

In order to assess throughput performance as a function capture threshold, Fig.5 shows throughput performance as a function SINR, for three different values of the capture threshold. Depending upon the channel quality as exemplified by the SINR on the abscissa in the figure, it could be preferable to operate at low capture threshold in order to gain higher throughput performance. The throughput predicted by Bianchi assuming SINR = ∞ and capture threshold = ∞, is depicted as a horizontal line along with the proposed model for comparison purpose. For decreasing capture threshold, the system throughput increases above the Bianchi's maximum achievable throughput performance. This is essentially due to the fact that, the capture effect tends to reduce the collision probability experienced by the contending stations which attempt simultaneous transmission.

Fig.6 shows the behavior of system throughput for basic access method as a function SINR, for two different payload sizes and for 5 transmitting stations. The upper curves are plotted for the payload size of 1024bytes and the bottom curves are plotted for the payload size of 128 bytes. In both curves, saturation throughput is depicted for two different values of capture threshold.

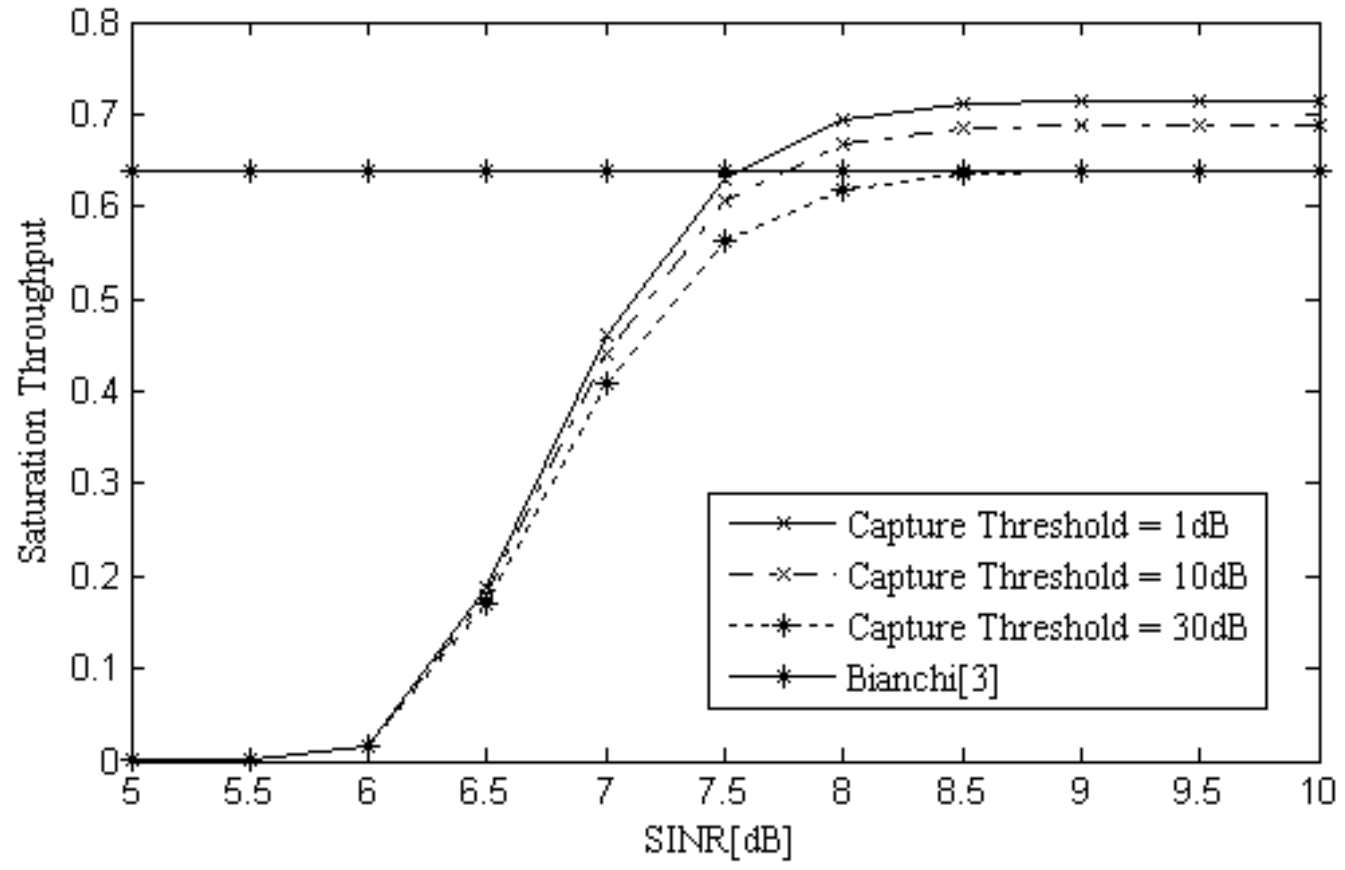

Figure 5. Saturation throughput for basic access mechanism as a function of SINR for capture thresholds 1dB, 10dB and 30dB, while payload size 1024 bytes and number of contending stations 5.

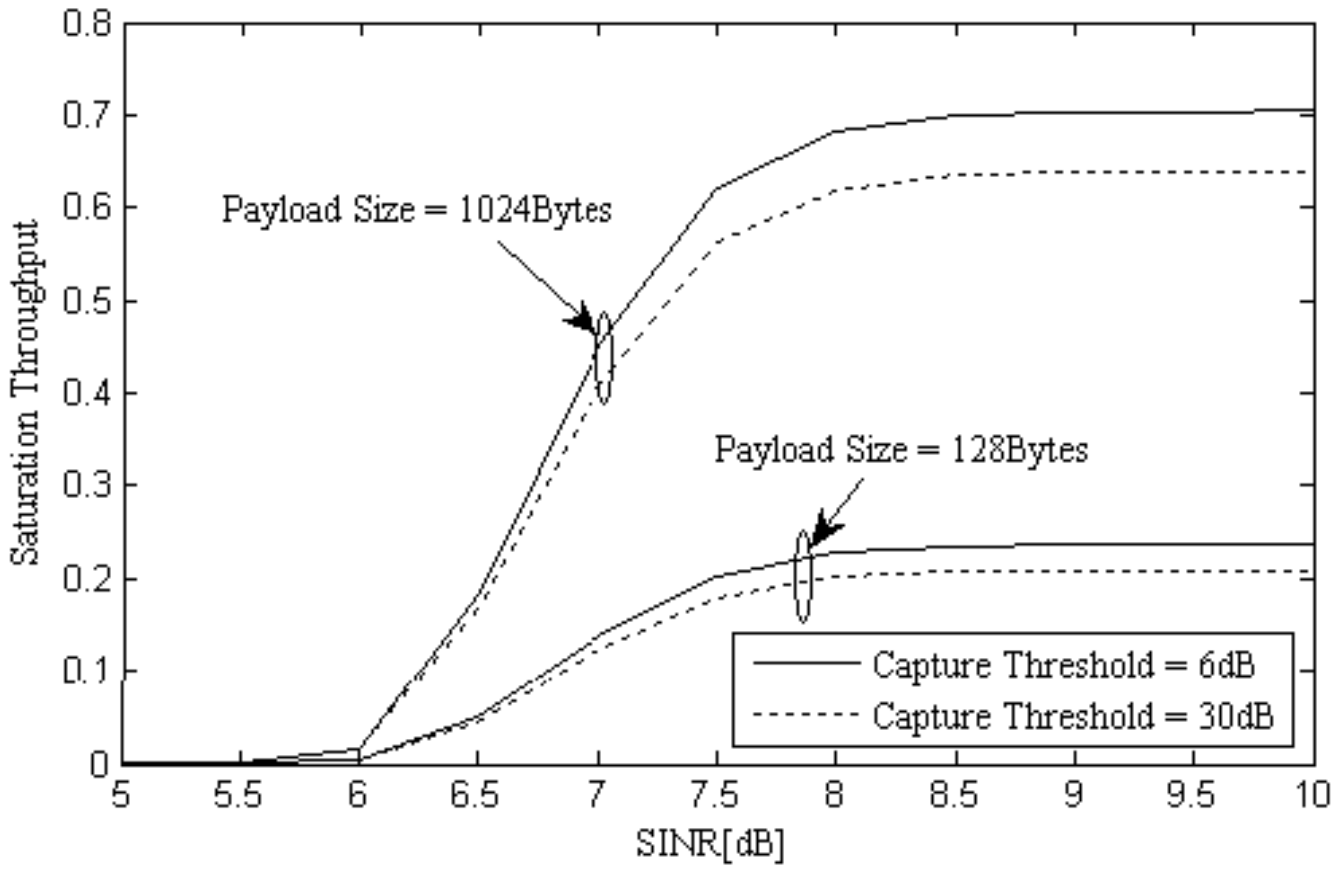

Figure 6. Saturation throughput for basic access mechanism as a function of SINR for payload sizes 1024 and 128 bytes, number of contending stations 5 and capture thresholds 6dB and 30dB.

Upon comparing the curves, it is easily seen that the system throughput performance is poor for low values of payload size. On the other hand, when the capture threshold is high, collision probability increases, that tend to reduce the throughput performance.

Fig.7 shows the behavior of saturation throughput for basic access method as a function SINR for two different capture thresholds. It can be easily noticed that, when channel errors are more, the achievable throughput is high due to proper rescheduling of contention window. On the other hand, for increasing capture threshold, throughput tends to reduce, as expected, in the presence of capture.

## IV. CONCLUSION

In this paper we have proposed a new MAC protocol for IEEE802.11 Distributed Coordination Function taking into account of both erroneous channel and capture effects. This avoids unnecessary idle slots by differentiating noise lost packets from collision lost packets, increasing throughput considerably. It performs as well as IEEE 802.11 in noisy environment considering low traffic conditions. Using the proposed model we have evaluated the throughput performance of IEEE802.11 DCF for basic access method. Based on this model we derive a novel and generalized expression for the station's transmission probability, which is more realistic, such as non ideal channel conditions. To the best of our knowledge, this paper is the first to show the undesirable behavior of the standard backoff procedure when transmission losses occur, to develop a practical solution to this problem, and to give a theoretical performance analysis under homogeneous link conditions.

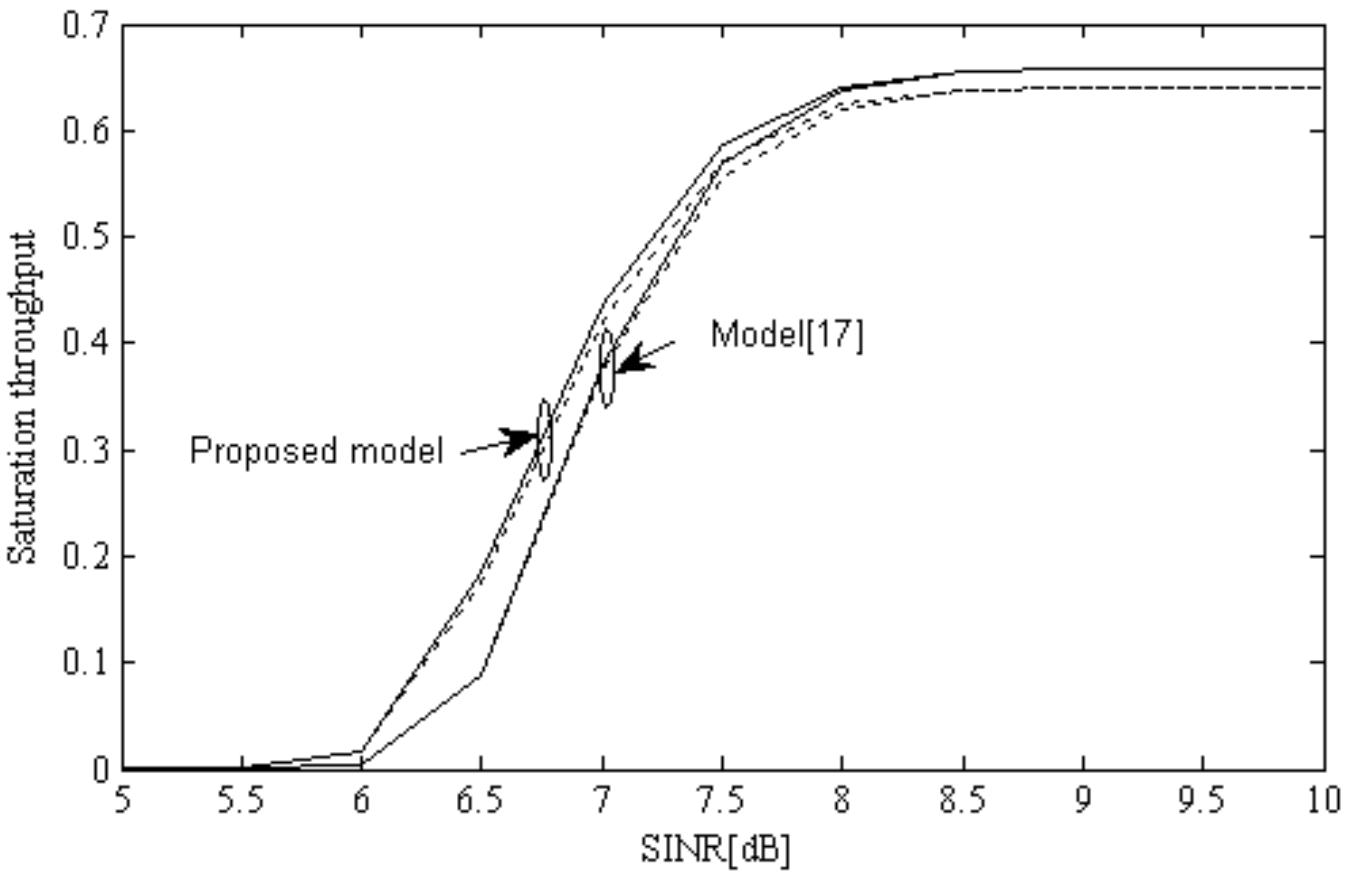

Figure 7. Saturation throughput for basic access mechanism as a function of SINR for capture thresholds 6dB and 24dB while payload size 1024 Bytes and number of contending stations 2.

## AUTHORS PROFILE

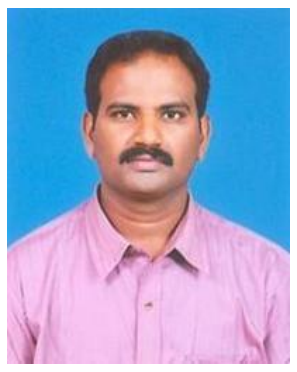

**Ponnusamy Kumar** received his B.E degree from Madras University, Chennai, India, in 1998, and the M.Tech degree from Sastra University, Tanjavore, India, in 2002. Since 2002 he is working as a Assistant Professor in K.S.Rangasamy College of Technology, Tamilnadu, India. He is currently pursuing his Ph.D degree under Anna University, Chennai, India. His research is in the general area of wireless communication with emphasis on adaptive protocols for packet radio networks, and mobile wireless communication systems and networks. Mr.P.Kumar is a member in IETE and ISTE.

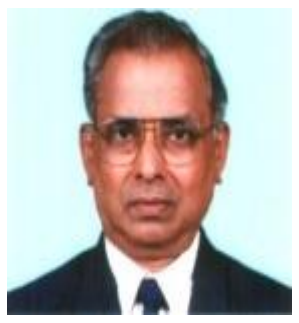

**A. Krishnan** received his Ph.D degree from IIT Kanpur, Kanpur, India. He is currently a professor with K.S.Rangasamy College of Technology, Tiruchengode, Tamilnadu, India. He has published over 150 papers in national and international journals and conferences. His research interests are in the area of communication networks, transportation, and hybrid systems. Dr. A. Krishnan is a senior member in IEEE, and member in IETE and ISTE.